# PEER-TO-PEER LIVE STREAMING AND VIDEO ON DEMAND DESIGN ISSUES AND ITS CHALLENGES

Hareesh.K[1]   and   Manjaiah D.H[2]

[1]Research Scholar, Jawaharlal Nehru Technological University, Anantapur, A.P, India
`mail_hareeshk@yahoo.com`
[2]Professor and Chairman, Department of CS, Mangalore University, Mangalore, India.
`ylm321@yahoo.co.in`

## ABSTRACT

*Peer-to-Peer Live streaming and Video on Demand is the most popular media applications over the Internet in recent years. These systems reduce the load on the server and provide a scalable content distribution. A new paradigm of P2P network collaborates to build large distributed video applications on existing networks .But, the problem of designing the system are at par with the P2P media streaming, live and Video on demand systems. Hence a comprehensive design comparison is needed to build such kind of system architecture. Therefore, in this paper we elaborately studied the traditional approaches for P2P streaming architectures, and its critical design issues, as well as practicable challenges. Thus, our studies in this paper clearly point the tangible design issues and its challenges, and other intangible issues for providing P2P VoD services.*

## 1. INTRODUCTION

With the advent of services and applications such as media streaming over the Internet, a client server service model is used.  But the speed and reliability have become critical issues for users. In client server model a client made a request to the video server, if the requested video is available in the video server, streamed to the requested client. One of the basic client service models is CDN (Content delivery Network) video streaming. In CDN approach a set of CDN server are placed at the network instead of downloading the video from the main multimedia server, a client normally directed to download the video from the nearby server (proxies). The CDN normally reduces the startup delays and also reduces the traffic over the network. The major challenge for server based video streaming requires scalability. A good quality of video stream requires high bandwidth. The major problem in video server that is CDN approach, as the number of client increases the bandwidth must proportionally increases. By sharing the load among various locations closer to the user-end, CDNs can deliver content to users in a timely manner. Content replication enhances robustness so that CDNs can maintain reliable service in case of failures. This makes server based video streaming solution more expensive.

To overcome the CDN approach problems, Peer to Peer networking has emerged as new paradigm and become a very successful in distributing the contents and also encourages the users to become both clients and servers which are commonly known as peers. In a P2P network, a peer not only download the data from the network but also upload the downloaded data to others users in a network. The uploading bandwidth of the client is utilized efficiently to reduce the load on the server. Unlike the client server system P2P system consists of peers interconnected with each other and self organized into network topologies in order to share the resources like bandwidth data, CPU cycles and buffer.

P2P Technologies were designed for specific file sharing applications such as Bit Torrent [2] and Emule [4.] In these file sharing applications, user need to wait until download the complete video in order to playback the video, it becomes longer delays on most of the cases due to intrinsic characteristics of the file and network bandwidth.

Streaming services over the Internet [16][18] system like Cool streaming [14] and PPLive [10] have been very successful in delivering Live media content over the internet to large number of end users. As a result P2P based VOD emerged a new challenge for the researchers and P2P technology. These systems are much more challenging to design and deploy than the Live streaming because a VOD server should allow the users to arrive at arbitrary times to watch media content or to use VCR like functionalities. In order to support VCR operations more efficiently, we believe that content- aware models which consider user-behavior according to the content being watched are necessary. Designing a P2P-VoD system with this kind of characteristics and information will improve overall system performance and reduce the load on the server.

The rest of the paper organized as follows: Section 2 describes the existing P2P live streaming systems Section 3 presents the P2P video-on-demand systems, within each section, representative systems are used as examples to show both tree-based and mesh-based system architectures. Section 4 and 5 describes design issues of Live and P2P VOD Systems; Section 6 shows other design challenges. Finally, the paper is concluded in Section 7.

## 2. P2P LIVE STREAMING

The video streaming services can be classified into two groups: Live streaming and video-on-demand streaming. In Live streaming allows video content to be transmitted in real time to all requesting users. One or more users have their playbacks synchronized to provide their stored content to other peers. On the other hand, video-on-demand users have the flexibility to watch any video at any moment in time, meaning that they do not need to synchronize their playback times. Moreover, they are capable to perform operations such as forward or backward on the media file. The following section gives a brief overview of the traditional overlay network structures for P2P live streaming systems.

### 2.1 Tree-based systems

The tree-based [17] P2P delivery has originated from IP multicast [7], where a single block is transmitted from the source and replicated by the routers along a distribution tree rooted at the source node. In *Cooperative Networking (Coop Net)* [12], for example, each node in the tree forwards the received stream to each of its children. This way, the load on the server is reduced by distributing content between cooperative users. The presence of a source node simplifies the task of locating content, since the root has all the essential information for constructing and maintaining the tree structure. Although an efficient solution for streaming audio and video, the use of computation and bandwidth resources along with the complexity of transport control for multicast sessions resulted in router overhead, causing IP level multicast to never appear entirely over the Internet. In this way, an implementation was done at the application layer, commonly known as Application-Level Multicast (ALM).

#### 2.1.1 Single-tree streaming

In single tree streaming, users taking part of a streaming session can get together to form an ALM tree having the source video server as the root. Figure 1 shows the tree structure with peers distributed at different levels receiving and forwarding information in a top-bottom direction. Overcast [8] and ESM [3] are considered pioneer works of single-tree based streaming. Both of them use a single ALM tree to distribute content; where each non-leaf peer in the tree has the function to retrieve and forward media content upon its arrival.

Figure 1 show Application Layer Multicast P2P video streaming .The figure also shows an ALM streaming tree with ten participating peers rooted at the source server node. The top level immediately below the root is formed by only two peers receiving media content directly from the source server. At the same time, these two peers push the content one level below, towards bottom, that includes four peers which receive the data, but only two of them forward it to end-leaf nodes located at the bottom level of the tree. Because leaf nodes usually account for a large portion of peers in the system, and they do not make any upload bandwidth contribution to it, the peer bandwidth utilization efficiency is affected to a great extent. Considering that peers at lower levels are the last ones to receive video content, the construction of a streaming tree with fewest levels possible is preferred;

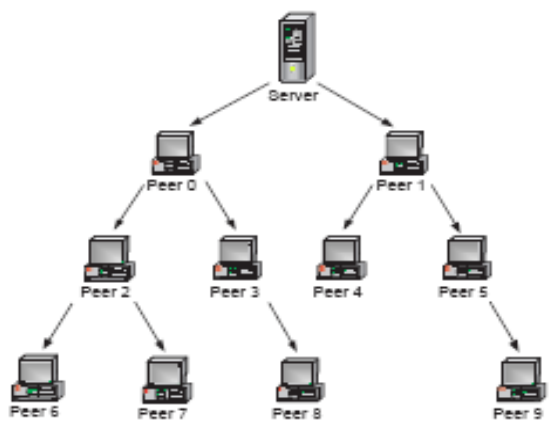

Figure1: Application Layer Multicast (ALM) P2P video Streaming

That is the tree topology should spread as much as possible at every level. Nonetheless, it is important to consider that a peer forming part of an internal node has uploading constraints which will determine its maximum fan out degree.

Maintaining this type of tree structures is also very important since users forming part of a P2P streaming session are known to be very dynamic. Users can join and leave the session at any moment. Therefore, the sudden departure of one of these peers, either gracefully or unexpectedly has machine crashes. It has a great impact in all its descendants since there is only one available path of streaming flow coming from the source. To handle this kind of disruption, the streaming tree needs to be recovered and recalculated by reassigning the affected nodes to the source server or other available peers. Unfortunately, for a large streaming system, the source server becomes the performance bottleneck and single point of failure. To address this problem, distributed algorithms such as ZIGZAG [13] have been developed constructing and maintaining streaming trees in a distributed manner. Despite of all the efforts, tree-based streaming has shown to be unable to recover fast enough to deal with frequent peer churn.

### 2.1.2 Multi-tree streaming

Multi-tree based techniques such as Bullet [9] have been proposed [20] to handle problems related to leaf nodes. In this approach the source server divides the stream into multiple sub-streams, as shown in Figure 2. A single streaming tree is split into various sub-trees, one for each sub-stream. Each peer joins all sub-trees available to retrieve sub-streams and has different positions in each one of them.

The upload bandwidth of a peer is utilized to upload a sub-stream whenever it is considered to be an internal node in some sub-tree. In order to have a fully balanced multi-tree streaming, a single peer is positioned on an internal node in only one sub-tree that is uploading one sub-stream from the level below, while positioned on a leaf node on the remaining sub-trees that is downloading a sub-stream from the level above.

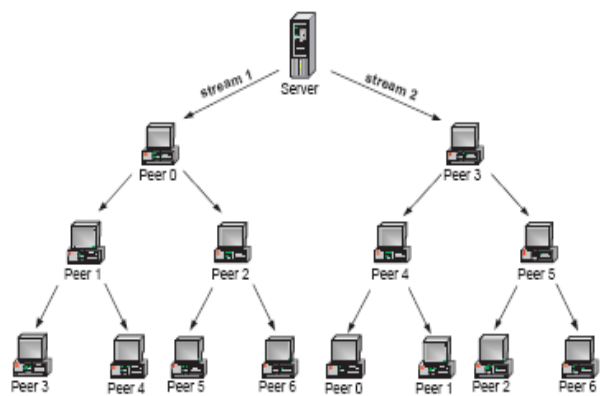

Figure 2 – Multi-tree based streaming

## 2.2. Mesh-based systems

Tree-based systems only allow peers to have one parent in a single streaming tree to download from; which causes a single point of failure. If the peer's parent leaves, the peer and all its children stop retrieving information until the peer connects to a different parent over again. To deal with this problem, many P2P streaming systems adopt a mesh-based approach [19][21],such as DONet/Coolstreaming [15]. The main characteristic of based streaming system is that there is no static streaming topology. At any given time, a peer collaborates with multiple neighboring peers, uploading/downloading video to/from multiple peers concurrently. Finding new peers to keep a desired level of connectivity is possible, and it is exactly what makes them particularly robust against peer churn. Analogous to P2P file sharing systems like BitTorrent [2], a mesh streaming system includes a tracker to keep track of peer activities during the video session. Every new peer wanting to join the session needs to contact the tracker and report its own information (i.e. IP address, port number). Afterward, the tracker provides a list of peers containing the information of a random subset of active peers available. Using this list, the peer attempts to initiate peering connections; and if successful, it starts exchanging video content with its neighbors. To handle unexpected peer departures, peers regularly exchange keep-alive messages. At the same time, depending upon system's peering strategies, a peer does not only connect to new neighbors in response to peer departures, but also when better streaming performance can be achieved.

In mesh-based systems, the concept of video stream becomes invalid due to the mesh topology. The basic data unit in this kind of systems is video chunk. The multimedia server divides the media content into small media chunks of a small time interval, each of them with a unique sequence number that serves as a sequence identifier. Later, each chunk is disseminated to all peers through the mesh. Since chunks may take different paths in order to reach a peer, they may arrive to destination in a non-sequential order. To deal with this matter, received chunks are normally buffered into memory and sequentially rearranged before delivering them to its media player, ensuring continuous playback. There are two different ways to exchange data in mesh-based systems: push and pull. In a mesh-push system, a peer constantly pushes a received chunk to the rest of neighbors who need it. However, due to system topology, redundant pushes could be responsible for wasting peer uploading bandwidth. As a result, a mesh-pull system is used to avoid the situation described earlier. This technique allows peers to periodically exchange chunk availability according to buffer maps. A buffer map holds the sequence number of chunks currently available in a peer's buffer; this way a peer can decide from which peers to download which chunks, avoiding redundant chunk transmissions that were present using push. A disadvantage of the pull technique is that both frequent buffer map exchanges and pull requests produce more signaling overhead and introduce additional delays while retrieving a chunk.

## 3. P2P VIDEO ON DEMAND

Video-on-demand (VoD) systems provide multimedia services offer more flexibility and convenience to users by allowing them to watch any kind of video at any point in time. Such systems are capable of delivering the requested information and responsible for providing continuous multimedia visualization. Compare to live streaming, in VoD systems the user has complete control over the media by making use of VCR operations such as pause, forward and backward functionalities (also known as jump operations); the same way as if the functionalities were used in a DVD player. VoD systems need to accommodate a large number of users watching the same video asynchronously, watching different parts of the same video at any given time. This is a very challenging design situation for tree-based P2P systems because users using this kind of overlay are synchronized and receive the content directly from the source server, and exactly in the same order it left the root node. Mesh-based P2P systems were successfully introduced to distribute large files and later applied to live streaming. In this kind of systems a large file is usually broken into many small blocks. Both the system throughput and the rate, at which the content can be distributed to users, greatly depend on the diversity of the blocks contained at different peers. The challenge of providing VoD services using mesh-based P2P networks lies in the fact that the blocks have to be received at the peer-side in a sequential order, and time constraints have to be considered at all times to guarantee continuous visualization. Therefore, supporting VoD services using mesh-based P2P is also a great effort. As we know that in tree-based and mesh-based P2P systems for live streaming have their own advantages and disadvantages. In this section, the different ways of how to adapt these two approaches to providing VoD services are described.

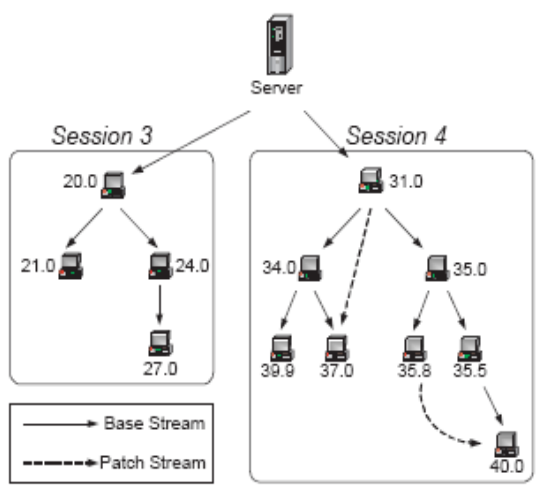

Figure 3 – Application Layer Multicast - P2Cast policy

## 3.1 Tree-based P2P VoD

The Patching [5] is one of the first IP multicast policies introduced for supporting VoD services. Inspired on this scheme, P2Cast [4] was designed for distributing video content among asynchronous users, where every user act as a server while retrieving media content. As shown in Figure 3, users arriving within a threshold form part of a session. Along with the source server, users belonging to the same session form an ALM tree, known as the base tree.

The entire video is then streamed from the source server using this base tree; just the same way as in tree-based P2P live streaming. Users joining the session join the base tree and retrieve the base stream from it. In contrast, new clients who missed the initial part of the video must obtain a patch directly from the server or other users who have already cached the required content. Users behave just like peers in a P2P network, and provide two main functions: *(i) base stream forwarding*, where users participating in the tree-based overlay should be capable of forwarding the received base stream to others; and *(ii) patch stream serving,* where users cache the initial part of the video and serve the patch to latecomers

## 3.2 Mesh-based P2P-VoD

In mesh-based P2P file sharing networks, a file is typically divided into a set of small size data blocks. The server, or also known as seeder, is in charge of distributing the blocks to different peers. Later, peers collect information of other users sharing the same content interests and form neighborhoods that allows them to exchange the blocks they are missing or willing to share. To maximize users upload bandwidth and consequently achieving the highest downloading throughput possible, block diversity needs to be taken into account at each one of the peers. Block diversity improves the systems throughput, but could become a problem during playback time since video blocks have to provide continuity and be played in sequential order. Users need to receive blocks sequentially and not in a random order to watch the movie while downloading [1]. Additionally, the nature of VoD systems demands the availability of different media blocks at any given time, especially if users decide to perform VCR operations during playback and expect high level of service with low startup delays in return.

## 4. DESIGN ISSUES OF P2P LIVE STREAMING

The following factors shows some of the common design issues of P2P live streaming.

### 4.1 End-to-end bandwidth is a key problem.

In a client server system, we can support a large number of users in a local area by adding more servers. For a global scaled service, simply adding servers is not enough however. The end-to-end bandwidth may limit the geographical distribution of the users. Hence, CDN is a possible solution, but it remains very expensive and is not readily deployable. On the other hand, P2P enables intelligent path selections that may avoid the above problem. Nonetheless, we also find that, comparing with client-server solution, the overlay solution may introduce additional delay for a user to smoothly playback the video.

### 4.2 Upload bandwidth is a physical limitation.

While P2P streaming is more flexible than client server, it does have certain limitations. The most significant is the demands on the upload bandwidth. For a successful P2P media streaming system, at least we should have average upload bandwidth larger than the streaming rate. Hence, ADSL and other asymmetrical Internet accesses become more common.

**4.3 ISP issues and traffic effects.**

Currently, a large portion of the available bandwidth at network edges and backbone links are occupied by such P2P applications as Bit Torrent [2]. Many ISPs thus limit this kind of traffic. For file sharing the limitation may only slow down the download, but for media streaming it can be fatal. Other filtering mechanisms may also create problem to P2P streaming systems.

## 5. DESIGN ISSUES IN P2P-VOD

P2P-VoD systems also deliver content by streaming, but unlike live streaming, peers are able to watch different parts of the video at the same time while Collaborating with each other and offloading the server. To make this collaboration possible, users need to contribute with a small amount of storage instead of just the playback buffer in memory. This section presents a general architecture and taxonomy for P2P-VoD systems, considering issues like service scheduling, replication strategies, and so on.

**5.1 Major System Components**

A P2P-VoD system has the following major components: a set of servers which act as source of content; a set of trackers in charge of helping peers connect to other peers to share the same content; a bootstrap server to help peers to find a suitable tracker (e.g. based on the geographical location), and other bootstrapping functions; other servers such as log servers used for data measurement purposes, and transit servers for helping peers located behind NAT boxes; finally, a set of peers running software downloaded from the P2P-VoD operator.

**5.2 Segment sizes**

Segmentation of content is fundamental in the design of P2P-VoD systems. There are some considerations to have in mind for making this decision. From the scheduling point of view, content should be divided into as many pieces as possible for flexibility reasons at the time of finding which piece to upload from which neighbor. From the overhead point of view, the larger the segment the better since it helps to minimize overheads. Another consideration is related to the real-time nature of streaming. The video player or set-top-box at the user-side expects a certain minimum size for a piece of content to be viewable and delivered according to some deadline, namely a chunk. If units are too large, the chances of not fulfilling this deadline increase.

**5.3 Replication Strategy**

Assuming each peer contributes with some amount of hard disc storage, a distributed P2P storage system is formed by the entire viewer population, each of them containing chunks. Once all pieces in a chunk are available locally, the chunk is advertised to other peers. The aim of replication strategy is to make chunks available to every user in the shortest time possible in order to meet with viewing demands. Design issues regarding replication strategies consider as: (i) allowing multiple movies to be cached if there is room on the hard disc. This is referred as multiple movie cache (MVC); and lets a peer watching a movie upload a different movie at the same time. (ii) to pre-fetch or not to pre-fetch; while pre-fetching could improve performance, it could also waste uplink bandwidth resources of the peer. (iii) Selecting which chunk or movie to remove when the disc cache is full; preferred choices for many caching algorithms are least recently used (LRU) or least frequent used (LFU).

### 5.4 Content Discovery

Together with a good replication strategy, peers must also be able to learn who is holding the content they need without introducing too much overhead into the system. P2P systems depend on the following methods for content discovery:

- a tracker; to keep track of which peers are replicating what part of the movie;
- DHT(Distributed hash table); used to assign movies to trackers for load balancing purposes;
- Gossiping method; used for chunk discovery even if the tracker is not available, a peer asks its neighbors for their chunk bitmaps and with this information, it selects which neighbor to download from.

### 5.5 Piece Selection

In order to download chunks from other collaborators, a peer uses a pull method and takes into account three important considerations for selecting which piece to download first: (i) sequential, selects the piece that is closest and necessary for playback; (ii) rarest first, selects the rarest piece in the system which in turn helps speeding up the proliferation of pieces; and (iii) anchor-based, fixed points defined to support VCR features (e.g. forward jumps), commonly found in VoD. Some approaches embed a hybrid strategy for piece selection, combining sequential and rarest first. Such is the case of Kangaroo, for example.

### 5.6 Transmission Strategy

Once a particular chunk has been selected for download, what happens if more than one neighbor has a copy of it? To answer this question, P2P-VoD systems rely on transmission strategy algorithms. This kind of algorithms are designed based on two objectives: (i) maximize download rate; and (ii) minimize the overhead caused by duplicate requests and transmissions

## 6. OTHER DESIGN CHALLENGES

P2P systems are designed to distribute the workload and network traffic among the peers and take advantage of the computing and storage resources of each individual peer. There are two aspects to this approach. One of the advantages of P2P system is very scalable and can potentially serve a very large streaming community where the network and processing load will be a significant challenge for a centralized system. The drawback of P2P systems is that because of the dynamic and unpredictable nature of peers a more complicated, fully distributed protocol is required to constantly maintain the system and recover from errors. A lack of centralized control also introduces difficulties for the administration and security of P2P systems. Below we list a few of the challenges commonly encountered in a P2P streaming system.

### 6.1 Quality of Service (QoS)

The quality of service (QoS) of a streaming system usually refers to the end-users experience. Criteria may include the smoothness of the display, the frequency of visual distortions, and the startup latency from session initiation to the onset of the display. QoS requirements depend on the type of P2P streaming systems. For example, users can tolerate a relatively longer delay in a non-interactive streaming system such as on-demand movie watching. This is in contrast to the requirements of a live, two-way audio conferencing system in which the delay must be bounded at the millisecond scale.

The fact that a P2P system is connected in a distributed topology introduces some challenges that are usually less relevant for a centralized system. For example, in order to accommodate a large number of members, P2P systems usually build an application-level overlay network among all users. The resulting stream forwarding or processing at the application level increases the end-to-end delay through the additional intermediate hops from source to destination and as a result it is difficult to build a low latency streaming platform using a P2P platform. One idea is to distinguish active users who require low latency from passive users who can tolerate longer latency. By clustering the active users logically close together the delay among them can be reduced .The remaining challenge is to distinguish active users effectively and automatically.

### 6.2 Dynamics

One of the major challenges for all P2P streaming systems is how to provide a reliable service over an unreliable, constantly changing and most likely, heterogeneous streaming architecture. The members of a P2P system are often of different computing power, network bandwidth and network connectivity. Some are connected from behind a firewall and some are connected through a network address translation (NAT) device. Peers may join and leave at any moment, leaving some fraction of the P2P streaming network isolated and disconnected. These dynamics make the construction of a reliable and deterministic streaming service become challenging task.

A common solution is to maintain redundant information to recover the lost service. This approach is easier to implement and the service may be more reliable since the server is usually monitored and maintained professionally. However such a centralized recovery design hinders the scalability of a distributed system and may increase its costs. A hybrid approach that combines the above two designs is often a good compromise

### 6.3 Security

Security is important for the P2P-VoD system to include mechanisms to authenticate content, so that the system is resistant to pollution attacks [11]. Such authentication can be implemented based on message digest or digital signature. Authentication can be done at two levels: chunk level or piece level. If chunk level, authentication is done only when a chunk is created and is stored to the hard disc. In this case, some pieces may be polluted and cause poor viewing experience locally at a peer. However, further pollution is stopped because the peer would detect a chunk is bad and discard it. The advantage of the chunk level authentication is its minimum overhead. Chunk-level authentication has at least two significant drawbacks. Sometime, polluted pieces may cause more damage than poor viewing experience, for example it may crash or freeze the player. Secondly, chunk is a rather large segment of content. There is some non-zero probability that apiece is bad not due to pollution; but this would cause the entire chunk to be discarded. For these reasons, it is better to do authentication at the piece level to ensure good video quality. In the current version of PPLive VoD, a weaker form of piece level authentication is also implemented, leveraging on the same key used for chunk level authentication.

## 7. CONCLUSION

In this paper, we studied the comparison of traditional approaches of P2P video streaming technology. We described several key P2P streaming designs, including system topologies, peering connections and data scheduling, that address various design challenges in providing large scale live and on-demand video streaming services on top of the best-effort Internet. Current deployments on the Internet demonstrate that P2P streaming systems are capable of streaming video to large users at lower cost and with minimal dedicated infrastructure. However, there are several fundamental limitations were observed in traditional approaches P2P video streaming solutions. In this paper, we also present a general architecture and important building blocks of realizing a P2P-VoD system. One can use this general framework and taxonomy to further study various design choices. The building blocks we described include the file segmentation strategy, replication strategy, content discovery and management, piece or chunk selection policy, transmission strategy and authentication. The P2P streaming designs issues are become most challenging task for the current researchers. This paper also provides the general framework for further research in P2P-VoD systems, in particular, to address the following important issues: how to design a highly scalable P2P-VoD system to support millions of simultaneous users; how to perform dynamic movie replication, replacement, and scheduling so as reduce the workload at the content servers; how to quantify various replication strategies so as to guarantee a high health index; how to select proper chunk and piece transmission strategies so as to improve the viewing quality etc.

**Authors**

Short Biography

Hareesh K[1]   is a Research Scholar, Jawaharlal Nehru Technological University, Anantapur, A.P, India. He received his BE in Computer Science and Engineering from Bangalore University and MTech in Computer Science and Engineering from Visvesvaraya Technological University, Belgaum, India. His area of interest includes Computer communication network, web Technology, mobile ad hoc networks, multimedia applications, database management systems, Internet technologies and quality of service. His current research includes P2P Video-on-Demand systems.

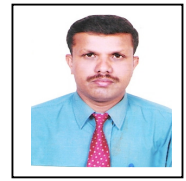

Manjaiah D.H.[2] is a Professor and Chairman in the Department of Computer Sciences, Mangalore University, and Mangalore, India. He received Ph.D degree from University of Mangalore, M.Tech. from NITK, Surathkal and B.E. from Mysore University, India. He has more than 15 years extensive academic, Industry and Research experience. He has authored more than 50 research papers in International / National reputed journals and conferences. His current research focuses on protocol validation and verification, network performance and analysis, Advanced Computer Networking, Mobile / Wireless Communication, Wireless Sensor Networks, Operations Research, E-commerce, Internet Technology and Web Programming.

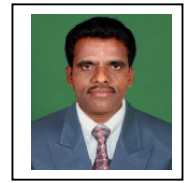